\documentclass[11pt]{article}

\usepackage{amsmath, amsthm}
\usepackage{newpxtext}
\usepackage{newpxmath}
\usepackage{complexity}
\usepackage{tikz}
\usepackage{float}
\usepackage{stackrel}

\usepackage{mathtools}
\usepackage{comment}

\usepackage{a4}

\usepackage{algorithm}
\usepackage{algpseudocode}
\usepackage{accents}

\usepackage{bm}

\usepackage{fancybox}
\usepackage{color}
\usepackage{latexsym}

\usepackage{eepic}
\usepackage{epic}
\usepackage{epsf}
\usepackage{graphics}
\usepackage{dsfont}

\newtheorem{theorem}{Theorem}
\newtheorem*{theorem*}{Theorem}

\newtheorem{lemma}[theorem]{Lemma}

\newtheorem{corollary}[theorem]{Corollary}
\newtheorem{claim}[theorem]{Claim}

\newtheorem{proposition}[theorem]{Proposition}
\theoremstyle{definition}

\newtheorem{definition}[theorem]{Definition}
\newtheorem{remark}[theorem]{Remark}

\newcommand{\B}{\mathfrak{B}}

\renewcommand{\C}{\mathfrak{C}}
\renewcommand{\M}{\mathcal{M}}

\renewcommand{\H}{\mathcal{H}}
\renewcommand{\E}{\mathcal{E}}
\renewcommand{\D}{\mathcal{D}}
\newcommand{\F}{\mathds{F}}

\RequirePackage[colorlinks=true]{hyperref}
\hypersetup{
  linkcolor=[rgb]{0.3,0.3,0.6},
  citecolor=[rgb]{0.2, 0.6, 0.2},
  urlcolor=[rgb]{0.6, 0.2, 0.2}
}

\renewcommand{\angle}[1]{\langle #1 \rangle}

\newcommand{\FX}{\mathds{F}\angle{X}}
\newcommand{\Perm}{\mathrm{Perm}}

\DeclareMathOperator{\supp}{\mathrm supp}

\DeclareFontFamily{OMX}{MnSymbolE}{}
\DeclareFontShape{OMX}{MnSymbolE}{m}{n}{
   <-6>  MnSymbolE5
   <6-7>  MnSymbolE6
   <7-8>  MnSymbolE7
   <8-9>  MnSymbolE8
   <9-10> MnSymbolE9
   <10-12> MnSymbolE10
   <12->   MnSymbolE12}{}
\DeclareSymbolFont{mnlargesymbols}{OMX}{MnSymbolE}{m}{n}
\SetSymbolFont{mnlargesymbols}{bold}{OMX}{MnSymbolE}{b}{n}
\DeclareMathDelimiter{\llangle}{\mathopen}{mnlargesymbols}{'164}{mnlargesymbols}{'164}
\DeclareMathDelimiter{\rrangle}{\mathclose}{mnlargesymbols}{'171}{mnlargesymbols}{'171}

\newcommand{\ncVP}{\mathsf{\VP_{nc}}}
\newcommand{\ncVNP}{\mathsf{\VNP_{nc}}}

\DeclareFontEncoding{LS1}{}{}
\DeclareFontSubstitution{LS1}{stix}{m}{n}
\DeclareSymbolFont{symbols2stix}{LS1}{stixfrak} {m} {n}
\DeclareMathSymbol{\lparenless}{\mathopen} {symbols2stix}{"32}
\DeclareMathSymbol{\rparengtr}{\mathclose}{symbols2stix}{"33}

\title{On Lifting Lower Bounds for Noncommutative Circuits using Automata}
\author{ V. Arvind\thanks{Institute of Mathematical Sciences (HBNI), and 
Chennai Mathematical Institute, Chennai, India. \texttt{Email: arvind@imsc.res.in.}} \and
Abhranil Chatterjee\thanks{Indian Statistical Institute, Kolkata, India.
    \texttt{Email: abhneil@gmail.com.} Research Supported by DST-INSPIRE Faculty Fellowship.} }

\date{\today}
\begin{document}
\maketitle

\begin{abstract}
  We revisit the main result of Carmosino et al \cite{CILM18} which
  shows that an $\Omega(n^{\omega/2+\epsilon})$ size noncommutative
  arithmetic circuit size lower bound (where $\omega$ is the matrix multiplication exponent) for a constant-degree
  $n$-variate polynomial family $(g_n)_n$, where each $g_n$ is a
  noncommutative polynomial, can be ``lifted'' to an exponential size
  circuit size lower bound for another polynomial family $(f_n)$
  obtained from $(g_n)$ by a lifting process. In this paper, we present a simpler and more conceptual automata-theoretic proof of their result. 
\end{abstract}

\section{Introduction}

Algebraic Complexity concerns itself with the complexity of algebraic
computations of multivariate polynomials. 
It starts with Strassen's work on matrix
multiplication from the 1960's. In the 1970's, Valiant defined the
algebraic complexity classes $\VP$ and $\VNP$~\cite{Val79}, which are analogues to
$\P$ and $\NP$, which brings to focus the problem of proving
superpolynomial arithmetic circuit size lower bounds for an explicit
polynomial family like the permanent $\Perm_n$ which is complete for
$\VNP$ under projection reductions. This research area
has a rich history, nicely described in the text by Burgisser et al
\cite{BCSbook}. It is believed that separating $\VP$ from $\VNP$ is easier than the
$\P$ vs $\NP$ problem. But the problem remains open despite intense
research and highly nontrivial progress in recent years \cite{LST,Neeraj} and the $\Omega(n\log n)$ circuit size lower bound
result of Baur and Strassen \cite{BS83} remains the best known lower bound to this date.

Nisan \cite{Ni91} initiated the study on the algebraic complexity of
\emph{noncommutative polynomials}. The noncommutative polynomial ring
$\FX$, where $X=\{x_1,x_2,\ldots,x_n\}$ is a set of $n$ free
noncommuting variables, consists of noncommutative polynomials which
are $\F$-linear combinations of words over $X$. Noncommutative
arithmetic circuits computing polynomials in $\FX$ are defined like
their commutative analogs. The only difference is that multiplication
gates in the circuit are not commutative. The classes $\ncVP$ and $\ncVNP$, 
which are noncommutative analogs of $\VP$ and $\VNP$, can be defined, as 
has been done by Hrubes et al \cite{HWY10b}. In the same article, it is 
shown that $\Perm_n$ is $\ncVNP$-complete under projections. The main 
lower bound question is to separate $\ncVP$ and $\ncVNP$, i.e.\ whether 
the noncommutative permanent $\Perm_n$ requires superpolynomial size 
noncommutative arithmetic circuits. Arguably, this question should 
be easier in the noncommutative case.  Indeed, Nisan \cite{Ni91} has 
shown an exponential lower bound on the size of a noncommutative formula 
(more generally, a noncommutative algebraic branching program) computing 
the noncommutative $\Perm_n$. However, it remains open for noncommutative 
circuits. Moreover, we do not have anything better than the $\Omega(n\log n)$ 
lower bound result of Baur and Strassen in the unrestricted setting. 
We note that, recently, Chatterjee and Hrube\v{s}~\cite{CH23} have obtained a quadratic 
lower bound for \emph{homogeneous} noncommutative circuits. 

Why is it so difficult to obtain even a quadratic lower bound for unrestricted
noncommutaive circuits? A few years ago, in 2018, Carmosino et al \cite{CILM18} showed that 
an $\Omega(n^{\omega/2+\epsilon})$ circuit size lower bound\footnote{Here $\omega$ is the
matrix multiplication exponent.} for a 
constant-degree $n$ variate polynomial family $(g_n)$ can be 
``lifted'' to an exponential circuit size lower bound for a 
polynomial family $(f_n)$ (which is obtained from $(g_n)$
by the lifting process). The Carmosino et al lifting result partly 
explains the lack of success in showing even superlinear (in the 
number of variables) circuit size lower bounds for explicit
polynomial families. The lifting result 
is reminiscent of Allender and Koucky's work in the Boolean circuit complexity
setting \cite{AK10}, where the authors exploit the self-reducibility
structure of some $\NC^1$-complete problems to show that a superlinear
$\TC^0$ circuit size lower bound for them can be lifted to
superpolynomial $\TC^0$ circuit size lower bound.

Before we present the contribution of this paper, it is worth mentioning a 
similar result due to Hrube\v{s}, Wigderson, and Yehudayoff~\cite{HWY10} which 
indeed predates \cite{CILM18}. They show that a super-linear 
lower bound on the \emph{width} of an explicit degree $4$ polynomial can be 
lifted to an exponential circuit size lower bound for an explicit 
noncommutative polynomial.

\paragraph*{This paper}

In this paper, we present a simple and a more structured automata-theoretic argument for the Carmosino 
et al result \cite{CILM18} stated above. In their paper, the main idea is to use an encoding scheme that reduces the number of variables exponentially incurring only a polynomial blow-up in the degree. The core of the argument is to show the following:
\begin{lemma}[Informal]
    A noncommutative circuit can be decoded efficiently.
\end{lemma}
In this paper, we prove this using ideas from algebraic automata theory. The main two 
ingredients of our proof are to show (a) an efficient representation of a decoder using 
a weighted automaton, and (b) the use of the Hadamard product to construct the decoded 
circuit. Our proof is not only short and simple but also conceptually more satisfying. 
We highlight two consequences for different choices of parameters (details in 
Section~\ref{lift-subsec}):
\begin{itemize}
\item Let $(g_N)$ be an explicit noncommutative p-family, where
$\deg(g_N)=t$ for some constant $t$ for each $N$, such that $\C(g_N)\geq \Omega(N^{\omega/2+\epsilon})$, 
where $\epsilon>0$ is a constant. Then there is an explicit p-family 
$(h_n)_n$ in $\ncVNP$ such that $(h_n)$ requires circuits of size $n^{\Omega(n)}$.

\item Suppose $(g_N)$ is an explicit noncommutative p-family, where each
$\deg(g_N)=(\log N)^{O(1)}$, and $g_N$ requires circuits of size 
$\omega(N^{\omega/2}\cdot \log N)$. Then there is an 
explicit p-family $(h_n)$ in $\ncVNP$ such that $\C(h_n)=n^{\omega(1)}$.
\end{itemize}

\section{Preliminaries}

We recall some algebraic complexity definitions for noncommutative
computation. 
Further details on
these definitions and basic results can be found in Nisan's seminal
paper \cite{Ni91}.

\begin{definition}[Noncommutative Arithmetic Circuit]
Let $\F$ be a field. A \emph{noncommutative arithmetic circuit} $C$
over $\F$ and noncommuting indeterminates $x_1,x_2,\ldots,x_n$ is a
directed acyclic graph (DAG) with each node of indegree zero labeled
by a variable or a scalar constant from $\F$: the indegree $0$ nodes
are the input nodes of the circuit. Internal nodes are gates of the
circuit, and are of indegree two.  They are labeled either by a $+$ or
a $\times$ (indicating the gate type). Furthermore, the two inputs to
each $\times$ gate are designated as left and right inputs prescribing
the order of gate gate multiplication. Each internal gate computes a
polynomial (by adding or multiplying its input polynomials), and the
polynomial computed at an input node is just its label.  A special
gate of $C$ is designated the \emph{output}. The polynomial computed
by the circuit $C$ is the polynomial computed at its output gate. An
arithmetic circuit is a \emph{formula} if the fan-out of every gate is
at most one. For a polynomial $f\in\F\angle{X}$ we denote by $\C(f)$ its optimal circuit size.\smallskip
\end{definition}



We recall some more definitions from Burgisser's text \cite{Burg,HWY10b,AJR18}.

\begin{definition}[p-family]
Let $\F$ be a field.  A sequence of multivariate noncommutative
polynomials $(f_n)$ over $\F$ is called a \emph{p-family} if there is
a polynomial $n^c$ that bounds both the degree and number of variables
in $f_n$ for each $n$. Suppose $f_n\in \F\angle{X_n}$ for each $n$.
The p-family $(f_n)$ is \emph{explicit} if there is a polynomial-time
algorithm that takes as input a monomial $m\in X_n^*$ and computes its
coefficient in $f_n$, for all $n$, and in time polynomial in $n$. For
example, the permament polynomial $(\Perm_n)_n$ is an explicit
p-family.
\end{definition}

\begin{remark}\label{explicit}
 In the definition of an explicit p-family, the running time of the
 algorithm that computes the coefficient of a monomial $m\in X_n^*$ is
 polynomial in the length of $m$ encoded in some \emph{fixed alphabet}
 like, for example, the binary alphabet. This point is important when
 we consider p-families --as indeed we will need to for the lower
 bound lifting result-- $(g_n)_n$ of constant degree polynomials where
 $\deg(g_n)\le t$ for $t$ independent of $n$.
\end{remark}

Some notation that we will use in this paper: for a polynomial
$f\in\F\angle{X}$ its support $\supp(f)=\{w\in X^*\mid$ coefficient of
$w$ is $\ne 0\}$ is the set of monomials with nonzero coefficient in
$f$. Thus, letting $f_w$ denote the coefficient of $w$ in $f$, we can
write $f=\sum_{w\in\supp(f)}f_w w$. 

\begin{definition}[Formal Power Series]
  Let $X$ be a set of free noncommuting variables and $\F$ be any
  field.  A \emph{formal power series} is a function $f:X^*\to \F$,
  where $X^*$ is the free monoid of all words (i.e. monomials) over
  $X$. We can equivalently denote the power series $f$ by the formal
  infinite sum $\sum_{w\in X^*}f(w) w$. The set of formal power series
  form a ring $\F\llangle X\rrangle$ over $\F$ known as the power
  series ring. Ring addition here is coefficient-wise and ring
  multiplication is the standard convolution product.
\end{definition}

We recall the definition of a weighted automata \cite{DK21}
with some basic details. Let $\mathcal{A}$ be a finite state automaton
with state set $Q$ with designated start state $s$ and final state
$t$. Let $R$ be any ring. Then $\mathcal{A}$ is an $R$-weighted
automaton if the transition function
\[
\delta: Q\times Y\times Q\to R
\]
assigns to every transition $(q_1,y,q_2)$ a weight $r_y\in
R$. Consequently, every monomial $w=y_1y_2\cdots y_r\in Y^*$ along an
$s$ to $t$ transition path $P$ in the automaton $\mathcal{A}$ is
assigned a weight $r_P\in R$ (which the product of the individual
weights for each transition step). The actual weight $r_w$ associated
with monomial $w$ is $r_w=\sum_P r_P$, where the sum is over all 
$s$ to $t$ transition paths $P$ for the monomial $w$ (and $r_w=0$ if
there are no such paths). We define the formal power series
\[
\sum_{w\in Y^*}r_w w
\]
to be the power series computed by the weighted automaton
$\mathcal{A}$. Equivalently, for each variable $y\in Y$ we have its
$|Q|\times |Q|$ state transition matrix $M_y\in
\mathcal{M}_{|Q|}(R)$. The $(i,j)^{th}$ entry of $M_y$ is the element
$\delta(i,y,j)\in R$. Then, corresponding monomial $w=y_1y_2\cdots
y_d\in Y^*$, the transition matrix is the matrix product
\[
M_w=\prod_{j=1}^d M_{y_j},
\]
and the coefficient $r_w$ of monomial $w$ in the power series computed
by $\mathcal{A}$ is the $(s,t)^{th}$ coefficient $M_w[s,t]$ of $M_w$.

\section{Lower Bounds via Efficient Decoding}

The proof of the lower bound lifting result \cite{CILM18} can be
described quite simply using some automata theoretic arguments.  It is
based on a simple encoder and decoder which can be described using
a weighted automata. We present the details in this section.

\subsection{Hadamard Product Computation}

The notion of Hadamard product is well-studied in algebraic automata
theory \cite[Theorem~5.5]{BR11}. It has also been used for
noncommutative polynomials to obtain some algebraic complexity results
\cite{AJS09,AMS10,AS10}.

For the purpose of this paper, we define the Hadamard product of a
noncommutative polynomial computed by a circuit and a formal
series computed by a small automaton.

\begin{definition}
  Let $f\in \F\angle{X}$ be a degree-$d$ polynomial and $S$ be a formal power
  series in $\F\llangle{X}\rrangle$, where $X$ is a finite set of free
  noncommuting variables. The \emph{Hadamard product} of $f$ and $S$
  is the noncommutative polynomial
\[
f \circ S = \sum_{m\in X^{\leq d}} [m]f\cdot [m]S\cdot m,
\]
where $[m]f$ and $[m]S$ denote the coefficients of the word $m$ in $f$
and in $S$, respectively.
\end{definition}







We recall the following result showing efficient Hadamard product
computation when the polynomial is computable by a small circuit and the series by a small automaton. 

\begin{theorem}{\rm\cite{AS18}}\label{matrix-valued}
Given a circuit $C$ and an automaton $B$ computing a homogeneous
degree-$k$ polynomial $f\in\F\angle{X}$ and a formal series $S\in
\F\llangle X \rrangle$ respectively, the Hadamard product polynomial
$f\circ S$ can be evaluated at any point $(a_1, a_2, \ldots,
a_n)\in\F^n$ by evaluating $C(a_1 M_1, a_2 M_2, \ldots, a_n M_n)$
where $M_1, M_2, \ldots, M_n$ are the transition matrices of $B$, and
the dimension of each $M_i$ is the size of $B$.
\end{theorem} 

If $C$ is given by black-box access then $(f\circ S)(a_1,\ldots,a_n)$ for $a_i\in\F, 1\le i\le n$ can be evaluated by evaluating $C$ on matrices defined by the automaton $B$~\cite{AS18} as follows: 
For each $i \in [n]$, the transition matrix $M_i$ in $\M_s(\F)$ are computed from the automaton $B$ (which 
is of size $s$) that encodes layers. We define $M_i[k,\ell] = [x_i] L_{k,\ell},$ where $L_{k,\ell}$ is the 
linear form on the edge $(k,\ell)$. Now to compute $(f\circ S) (a_1,a_2,\ldots,a_n)$ where $a_i \in \F$ for 
each $1\leq i \leq n$, we compute $C(a_1 M_1,a_2 M_2,\ldots a_n M_n)$.  The value $(f \circ S) (a_1,a_2,\ldots,a_n)$ is the $(1,s)^{th}$ entry of the matrix $f(a_1 M_1,a_2 M_2,\ldots, a_n M_n)$.


Theorem~\ref{matrix-valued} can be used to efficiently compute a
circuit for the Hadamard product polynomial $f\circ S$. Replace each
$x_i$ by $y_ix_i$ in the automaton $B$. Let $M_1, \ldots, M_n$ in be
the transition matrices where each entry is a linear form in $Y$
variables. We can now compute $f\circ S$ by evaluating $C(M_1, \ldots,
M_n)$ on the matrices $M_i, 1\le i\le n$. In this evaluation each
multiplication gate of the circuit $C$ actually denotes matrix
multiplication.  Hence we have the following.

\begin{theorem}\label{abp-circuit}
  Given a noncommutative circuit of size $s'$ computing a degree $k$
  polynomial $f \in \F\angle{X}$ and an automaton of size $s$
  computing a formal series $S \in \F\llangle{X}\rrangle$, we can
  compute a noncommutative circuit of size $s's^{\omega}$ for the
  noncommutative polynomial $f \circ S$ in deterministic time
  $s's^{\omega}\cdot \poly(n, k)$, where $\omega$ denotes the matrix
  multiplication exponent.\footnote{The current best algorithm for
  matrix multiplication, which is due to Alman and Williams \cite{AW21},
  shows $\omega<2.373$.}
\end{theorem}

\subsection{An Efficient Decoder using Weighted Automata}

We first define the encoding scheme. Let $X = \{x_0, x_1, \ldots,
x_{n-1}\}$, $Y = \{y_0, y_1, \ldots, y_{m-1}\}$ be disjoint sets of
noncommuting variables and let $X^*$ and $Y^*$ denote the free monoids
of words/monomials in $X$ and $Y$, respectively.

A \emph{monoid homomorphism} is a mapping
\[
h:X^*\to Y^*
\]
such that $h(\epsilon)=\epsilon$ and $h(ww')=h(w)h(w')$, where we
denote the empty word universally by $\epsilon$.

A mapping $h:X\to Y^*$ is \emph{prefix-free}if for any $x,x'\in X$
$h(x)$ is not a proper prefix of $h(x')$. Any such prefix-free mapping
$h$ can be uniquely extended to an injective monoid homomorphism
$h:X^*\to Y^*$, and we refer to it as an \emph{encoder}.  We will
first consider the following simple encoder.

\begin{definition}[Encoder]\label{enc-def}
Let $X = \{x_0, x_1, \ldots, x_{n-1}\}$, $Y = \{y_0, y_1, \ldots,
y_{m-1}\}$ be disjoint sets of noncommuting variables where $n = m^3$.
For each $i\in \{0, 1, \ldots, n-1\}$ let $j_i k_i \ell_i$ denote the
base-$m$ representation of $i$, where each $j_i,k_i,\ell_i\in
\{0,1,\ldots,m-1\}$. The encoder is the monoid homomorphism $\E:
X^*\to Y^*$ that uniquely extends the substitution map
$\E(x_i)=y_{j_i}y_{k_i}y_{\ell_i}$.
\end{definition}

The encoder $\E:X^*\to Y^*$ of Definition~\ref{enc-def} naturally
extends by linearity to polynomials. Thus, $\E:\FX\to \F\angle{Y}$
encodes noncommutative polynomials in $X$ into noncommutative
polynomials in $Y$.

\subsection*{The decoder automaton}

A decoder $\D:Y^*\to X^*$ is a map such that $\D(\E(m))=m$ for all
monomials $m\in Y^*$. By linearity, for any polynomial $h\in \FX$ we
have $\D(\E(h))=h$.

As summarized in the following lemma,, it is convenient to formally use weighted automata to describe the
decoder corresponding to $\E$. 
Let the ring $R$ be the free noncommutative
polynomial ring $\F\angle{X}$. Assume that the elements of
$\F\angle{X}$ commute with variables in $Y$. Then the formal series
which defines the decoder $\D$ is $\sum_{u\in X^*}u\E(u)$. Notice that in
this formal series, for $w=\E(u)$ we have $r_w=u$ and $r_w=0$
for all $w\in Y^*$ not in the range of the encoder $\E$.

\begin{lemma}\label{dec-aut-lem}
The series $S = \sum_{w\in X^*} w \E(w)\in \F\angle{X}\llangle Y
\rrangle$ is computable by an $\F\angle{X}$-weighted automaton of size
$2(m+1)$, which is the decoder $\D$ corresponding to the encoder $\E$,
and $m=|Y|$.
\end{lemma}

\begin{proof}
As $xy=yx$ for all $x\in X$ and $y\in Y$, we observe that the power series 
$S=\sum_{u\in X^*}u\E(u)$ has the following simple expression:
\[
S = \left(\sum_{i=1}^n x_i \E(x_i)\right)^*.
\]

Now, consider the following automaton $A$ of size $2m+2$ (see Figure~\ref{fig1}).

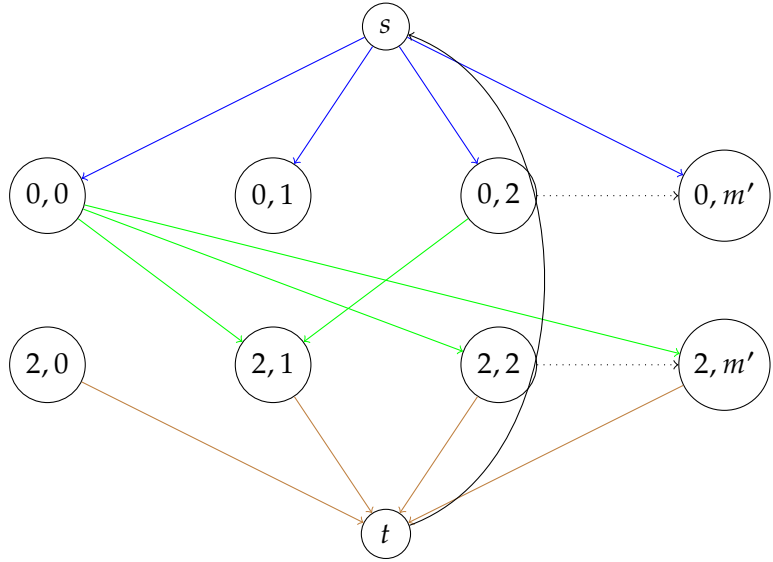
\begin{figure}[h]
\begin{center}
\begin{tikzpicture}[scale=0.75]
\node(pseudo) at (-1,0){};
\node(1) at (2.5,4)[shape=circle,draw]        {$s$};
\node(2) at (-3.5,1)[shape=circle,draw]        {$0,0$};
\node(3) at (.5,1)[shape=circle,draw] {$0,1$};
\node(4) at (4.5,1)[shape=circle,draw] {$0,2$};
\node(5) at (8.5,1)[shape=circle,draw] {$0,m'$};
\node(6) at (-3.5,-2)[shape=circle,draw]        {$2,0$};
\node(7) at (.5,-2)[shape=circle,draw] {$2,1$};
\node(8) at (4.5,-2)[shape=circle,draw] {$2,2$};
\node(9) at (8.5,-2)[shape=circle,draw] {$2,m'$};
\node(10) at (2.5,-5)[shape=circle,draw]        {$t$};

\path [->]


  
   (1)      edge [bend right=0] [color = blue] node [below]  {}     (2)
    (1)      edge [bend right=0] [color = blue] node [below]  {}     (3)
      (1)      edge [bend right=0] [color = blue] node [below]  {}     (4)
      (1)      edge [bend right=-0] [color = blue] node [below]  {}     (5)

        (6)      edge [bend right=0] [color = brown] node [below]  {}     (10)
    (7)      edge [bend right=0] [color = brown] node [below]  {}     (10)
      (8)      edge [bend right=0] [color = brown] node [below]  {}     (10)
      (9)      edge [bend right=0] [color = brown] node [below]  {}     (10)
      
              (2)      edge [bend right=0] [color = green] node [below] {} (7) 
    (2)      edge [bend right=0] [color = green] node [below] {} (8)
      (2)      edge [bend right=0] [color = green] node [below] {} (9) 
      (4)      edge [bend right=0] [color = green] node [below] {} (7) 
      
      (4) edge[dotted] (5)
	(8) edge[dotted] (9)
	(10) edge[bend right=70] node [above] {} (1) ; 
\end{tikzpicture}
\caption{The transition diagram of the automaton $A$}\label{fig1}
\end{center} 
\end{figure}

We describe the automaton in some detail because in Section~\ref{disc}
we will discuss this further.
The automaton has four layers. The initial layer has just the start
state $s$. The second and third layers each have $m$ states. The
final layer has just the final state $t$ from which the automaton
loops back to the start state $s$ on an $\epsilon$-transition\footnote{Strictly 
speaking we should remove the $\epsilon$-transition and directly go to state $(0,j)$ in
the second layer on reading $y_j$}. 

We now describe the role of the states in the second and third layers of 
the automaton.

Let $T = \{0,1, \ldots, m-1\}$. For each $j\in T$, we define a
transition from state $s$ to state $(0,j)$ reading $y_j$ (the state
$(0,j)$ encodes the symbol $y_j$ it has seen previously) and $(2,j)$
to $t$ reading $y_j$ (the state $(2,j)$ encodes the symbol $y_j$ it
will see next). 

The transitions between the second and third layers is where the
decoding actually happens. Between any pair of states $(0,i)$ in
the second layer and $(2,j)$ in the third layer, $i,j\in T$, 
the automaton has a weighted transition on input $y_k, k\in T$ which has 
weight $x_{\sigma(i, j, k)}$, where
$\sigma: \{0,1,\ldots,m-1\}^3\to \{0,1,\ldots,n-1\}$ is the bijection
\[
\sigma(i, j, k) = m^2i+m k + j.
\]

Notice that between $(0,i)$ and $(2,j)$ we have $m$ transitions, one
for each $y_k, k\in T$. The simple information-theoretic idea in this
construction is that the states $(0,i)$, $(2,j)$ and the transition
on $y_k$ hold the complete information about the string $y_iy_jy_k$
which the decoder can substitute with $x_{\sigma(i,j,k)}$. 
\end{proof}

\begin{remark}
  We refer to the above encoder as the $1$-to-$3$ encoder. In
  Section~\ref{disc}, where we discuss possibilities of improvements
  to the lower bound lifting result, we will consider the more general
  $1$-to-$r$ encoder.
\end{remark}

\subsection{The Lower Bound Lifting Result}\label{lift-subsec}

We are now ready to present the automata-theoretic proof of
the lower bound lifting result of \cite{CILM18}: namely,
that a circuit size lower bound of $\Omega(n^{\omega/2+\epsilon})$ 
for an explicit p-family $(g_n)$ \emph{of degree-$t$ polynomials} 
can be ``lifted'' to obtain an exponential circuit size lower bound 
for an explicit p-family $(h_n)$. Notice that the definition of 
explicit p-families applies to the constant-degree p-family 
$(g_n)$ in the sense explained in Remark~\ref{explicit}.

The result is an easy consequence of Theorem~\ref{abp-circuit}. In
fact we will show stronger result, as the simple analysis in the proof 
goes through for the choice of $t=O(\log n)$ and $\epsilon=O(\log\log n/\log n)$.
This yields the two consequences stated in the abstract.

We begin with showing that the decoder $\D$ preserves circuit size 
quite efficiently.

\begin{lemma}[efficient decoding]\label{lemma:decoding}
For a noncommutative polynomial $h\in\F\angle{X}$ suppose its encoding
$\E(h)\in\F\angle{Y}$ has a noncommutative circuit of size $s$. Then
$h$ has a noncommutative circuit of size bounded by $m^\omega\cdot s$,
where $m=|Y|$. More precisely, 
\[
\C(h)\le O(m^\omega)\cdot \C(\E(h)).
\]
\end{lemma}

\begin{proof}
The idea is to use the weighted automaton of Lemma~\ref{dec-aut-lem}
which defines the decoder $\D$ which computes the formal series
$S$. We first observe the following easy claim, that the Hadamard 
product $\E(h)\circ S$ evaluated at $y_j=1, 0\le j\le m-1$ is precisely
$h(X)$.
\begin{claim}
$h(X) = (\E(h)\circ S)(1, 1, \ldots, 1)$.
\end{claim}
Writing $h=\sum_{w\in\supp(h)} h_w\cdot w$, notice that we have
$\E(h)=\sum_{w\in\supp(h)} h_w\cdot \E(w)$. Thus we have
\[
\E(h)\circ S = \sum_{w\in\supp(h)} h_w\cdot w\cdot \E(w),
\]
noting that we are considering $S$ as a formal series in the $Y$ 
variables with coefficients as polynomials in the $X$ variables. 
Thus, the evaluation of $\E(h)\circ S$ for $Y$ variables substituted 
with $1$ will yield $h=\sum_{w\in\supp(h)} h_w\cdot w$. This proves
the claim.

As the size of the decoder automaton in Lemma~\ref{dec-aut-lem} is $2m+2$, 
the proof of the lemma follows from Theorem~\ref{abp-circuit} which
gives the claimed bound on the circuit size of the Hadamard product 
of a circuit with a weighted automaton.
\end{proof}


\begin{theorem}\label{cilm-lb-thm}
Let $(g_n)_n$ be an explicit noncommutative p-family, where
$\deg(g_n)=t$ for some constant $t$ for each $n$, such that $\C(g_n)\geq \Omega(n^{\omega/2+\epsilon})$, 
where $\epsilon>0$ is a constant. Then there is an explicit p-family 
$(h_n)_n$ in $\ncVNP$ where $h_n$ is $n$-variate with $\deg(h_n)=\poly(n)$ 
such that $\C(h_n)=n^{\Omega(n)}$.
\end{theorem}

\begin{proof}
Set $d=\lceil\log_3n\rceil$ and $N=n^{3^d}$. By assumption we have $\C(g_N)=\Omega(N^{\omega/2+\epsilon})$, where $\deg(g_N)=t$. By a 
$d$-fold application of the encoder $\E$ to the polynomial $g_N$, we 
obtain the polynomial 
\[
h_n =\E^d(g_N),
\]
where $h_n\in\F\angle{Y_d}$, letting $Y_d$ denote the set of
noncommuting variables in the output polynomial produced by $d$ 
applications of the encoder $\E$.

In general, for $1\le k\le d$ notice that $\E^k(g_N)\in\F\angle{Y_k}$, 
where $Y_k$ is a set of $N_k=n^{3^{d-k}}$ noncommuting variables, and
the degree of $\E^k(g_N)$ is $t\cdot 3^k$. Notice that 
$N_{k+1}^3=N_k$ for each $k\ge 1$ and $|Y_d|=N_d=n$. Therefore, 
$h_n(Y_d)$ is an $n$-variate polynomial of degree precisely $t3^d=tn$.

\begin{claim}
$\C(h_n)=n^{\Omega(n)}$.
\end{claim}

We will prove the claim by an inductive argument. More precisely, 
note that $\E^0(g_N)=g_N$ and $\E^d(g_N)=h_n$. Let 
$n_k=\epsilon(N)\cdot 3^k, 0\le k\le d$. By assumption, we have
$\C(\E^0(g_N))=\C(g_N)=\Omega(N^{\omega/2+\epsilon(N)})=\Omega(N^{\omega/2+n_0})$.

Suppose, as induction hypothesis that $\C(\E^k(g)) = \Omega (N^{\omega/2 + n_k}_k)$.
Then, by Lemma~\ref{lemma:decoding} we have 
\[
\C(\E^k(g_N))\le \alpha\cdot\C(\E(\E^k(g_N)))\cdot N_{k+1}^{\omega},
\]
for some constant $\alpha>1$. That implies   
\[
\C(\E^{k+1}(g_N)) \geq \frac{\alpha N_k^{\omega/2 + n_k}}{N_{k+1}^{\omega}}
= \frac{\alpha N_k^{\omega/2 + n_k}}{N_k^{\omega/3}} = \alpha N_{k+1}^{\omega/2 + n_{k+1}}.
\]

Putting it together, therefore, $h_n=\E^d(g_N)$ is $n$-variate in the variables $Y_d$ 
of degree $t3^d=t\cdot n=\poly(n)$ and 
\[
\C(h_n)=\C(\E^d(g_N)) = \Omega(n^{\omega/2 + 3^d\epsilon})=n^{\Omega(n)}.
\]
This completes the proof.
\end{proof}

In the above proof, if we let $t$ be a function of $N$, notice that 
choosing $t(N)=(\log N)^c$ with other parameters remaining the same, 
still guarantees $(h_n)_n$ to be an explicit p-family with $\deg(h_n)=\poly(n)$ 
and the lower bound  holds for $\C(h_n)$ as well. Furthermore, suppose 
we allow $\epsilon$ to be a variable quantity and set
$\epsilon=\omega\left(\frac{\log\log N}{\log N}\right)$.\footnote{Here $\omega(\cdot)$ is
the standard asymptotic notation and not the matrix multiplication exponent.} Then the lower bound 
assumption becomes 
\[
\C(g_N)=\Omega(N^{\omega/2+\epsilon(N)}) = \omega(N^{\omega/2}\cdot \log N),
\]
where $g_N$ is of degree $(\log N)^c$. In particular, this assumption is weaker
than that of Theorem~\ref{cilm-lb-thm}. Following the analysis in the proof
of Theorem~\ref{cilm-lb-thm} we obtain the following

\begin{corollary}\label{lb-cor}
Let $(g_N)_N$ be an explicit noncommutative p-family, where
$\deg(g_N)=(\log N)^c$ for constant $c>0$ and each $n$, such 
that $\C(g_N) = \omega(N^{\omega/2}\cdot \log N)$. Then there is an 
explicit p-family $(h_n)_n$ in $\ncVNP$ where $h_n$ is $n$-variate with 
$\deg(h_n)=\poly(n)$ such that $\C(h_n)=n^{\omega(1)}$.
\end{corollary}

\section{Discussion}\label{disc}

Can this lower lifting result be improved? As noted in \cite{CILM18}, the 
hardness assumption becomes $\C(g_N)=N^{1+\epsilon}$ if the matrix multiplication exponent $\omega=2$. Furthermore, the hardness assumption in Corollary~\ref{lb-cor} becomes $\omega(N\log N)$
for a degree $(\log N)^{O(1)}$ polynomial. Baur and Strassen's lower bound is $\Omega(N\log d)$ 
for an explicit degree-$d$ $N$-variate polynomial. Compared to that the $\omega(N\log N)$ lower
bound assumption translates
to $\omega(Nd^{\alpha})$ for some $\alpha>0$. Can the degree bound of $(\log N)^{O(1)}$ be 
relaxed in Corollary~\ref{lb-cor}?

We crucially use the Hadamard product construction described in Lemma~\ref{abp-circuit},
for which the circuit upper bound is $O(s'^\omega s)$ where $s'$ and $s$ are the given automaton and circuit 
sizes respectively. Matrix multiplication is inherent here. For, suppose there was a Hadamard
product construction with circuit upper bound $O(s'^\alpha s^\beta)$. Now, we can easily reduce
the multiplication of two $s'\times s'$ matrices to the Hadamard product of an automaton of size $O(s')$
and a circuit of size $s=O(1)$. Hence, it follows that $\alpha=\omega$. 

Another place where there is arguably some room for improvement is in the choice of the
encoder function and decoder automaton construction (Lemma~\ref{dec-aut-lem}). We note
that the decoder automaton of size $2m+2$ for the $1$-to-$3$ decoder is already optimal
to a constant factor. This is because we cannot have a $o(m)$ size automaton for $\D$
due to simple information-theoretic reasons. To see this, we observe that the decoder has to
output a variable $x_{\sigma(i,j,k)}\in X$ on a single transition edge, call it $e=(s_1,s_2)$.
But that means the information in the states $s_1, s_2$ and the input read on the transition
must contain the complete information about the triple $(i,j,k)$, where 
$i,j,k\in\{0,1,\ldots,m-1\}$ which is impossible if there are only $o(m)$ many states as
the number of edges need to be $\Omega(m^2)$.

\paragraph{The one-shot decoder and directly lifted lower bound}

Finally, we note that instead of using $1$-to-$3$ decoder $d$ times we
can directly decode $\E^d$ which uniquely encodes each $x_i, 1\le i\le N=n^{3^d}$ 
into a string in $Y^{3^d}$, where $Y=\{y_1,y_2,\ldots,y_n\}$. 
Let $\D^d$ denote the corresponding decoder. An automaton for $\D^d$ of
size $2n^{(3^d-1)/2}+2$ can be constructed exactly on the same lines
as Lemma~\ref{dec-aut-lem}. The automaton has four layers. The first
has the start state $s$ and the last has the final state $t$. The
second and the third layers have $n^{(3^d-1)/2}$ states each. From
the start state the automaton reads a prefix of length $(3^d-1)/2$
and remembers it in the state $s_1$ that it reaches in the second
layer. Likewise, each state $s_2$ in the third layer corresponds
to a suffix of length $(3^d-1)/2$. The transition $(s_1,s_2)$
reads the middle letter which, together with $s_1$ and $s_2$, describes the entire word over $Y$
of length $3^d$. This automaton has $M=2n^{(3^d-1)/2}+2$ states.
Now, applying Lemma~\ref{abp-circuit} we get
\[
\C(g_N)\le O(M^\omega)\cdot \C(\E^d(g_N))=O(M^\omega)\cdot \C(h_n).
\]
As $N=n^{3^d}$, by substituting we obtain $
\C(h_n) \ge n^{3^d\epsilon+\omega/2}=n^{\Omega(n)}$
for constant $\epsilon$, which proves Theorem~\ref{cilm-lb-thm}.


\end{document}